\documentclass[twoside,version1996]{siggraph-2000}
\usepackage{graphicx}
\usepackage{fancyhdr}
\usepackage{xcolor}
\usepackage{hyperref}

\suppresscover

\begin{document}


\newlength{\basebaselineskip}
\setlength{\basebaselineskip}{\baselineskip}

\newlength{\textbaselineskip}
\setlength{\textbaselineskip}{0.98 \basebaselineskip}

\newlength{\captbaselineskip}
\setlength{\captbaselineskip}{0.90 \basebaselineskip}

\newcommand{\BM}[1]{
  \mbox{\boldmath$#1$}
}
\newcommand{\RM}[1]{
  \mbox{$\mathrm{#1}$}
}

 

\acmcategory{research}
\acmformat{print}

\contactname{Jessica K. Hodgins}
\contactaddress{College of Computing\\
   801 Atlantic Drive \\
   Georgia Institute of Technology \\
   Atlanta, GA 30332-0280 }
\contactphone{(404) 894-9763}
\contactfax{(404) 894-0673}
\contactemail{jkh@cc.gatech.edu}

\estpages{8}



\title{Animating Explosions}
\author{
  Gary D. Yngve \and James F. O'Brien \and Jessica K. Hodgins
}
\affiliation{
  GVU Center and College of Computing\\
  Georgia Institute of Technology
}

\maketitle

\setlength{\baselineskip}{2in}
\setlength{\baselineskip}{\captbaselineskip}
\begin{abstract}
  In this paper, we introduce techniques for animating explosions and
  their effects.  The primary effect of an explosion is a disturbance
  that causes a shock wave to propagate through the surrounding
  medium.  This disturbance determines the behavior of nearly all
  other secondary effects seen in explosions.  We simulate the
  propagation of an explosion through the surrounding air using a
  computational fluid dynamics model based on the equations for
  compressible, viscous flow.  To model the numerically stable
  formation of shocks along blast wave fronts, we employ an
  integration method that can handle steep pressure gradients without
  introducing inappropriate damping.  The system includes two-way
  coupling between solid objects and surrounding fluid.  Using this
  technique, we can generate a variety of effects including shaped
  explosive charges, a projectile propelled from a chamber by an
  explosion, and objects damaged by a blast.  With appropriate
  rendering techniques, our explosion model can be used to create such
  visual effects as fireballs, dust clouds, and the refraction of
  light caused by a blast wave.
\end{abstract}

\setlength{\baselineskip}{\captbaselineskip}
\begin{CRcatlist}
  \CRcat{I.3.5}{Computer Graphics}{Computational Geometry and Object Modeling}{Physically based modeling};
  \CRcat{I.3.7}{Computer Graphics}{Three-Dimensional Graphics and Realism}{Animation};
  \CRcat{I.6.8}{Simulation and Modeling}{Types of Simulation}{Animation}
\end{CRcatlist}

\setlength{\baselineskip}{\captbaselineskip}
\keywordlist
\setlength{\baselineskip}{\textbaselineskip}



\setcounter{page}{29}


\renewcommand{\footrulewidth}{0pt}
\thispagestyle{fancy}
\fancyfoot{}
\fancyhead{}
\fancyfoot[RO,LE]{\thepage}
\fancyhead[LE]{\fontsize{8}{8}\textsf{ACM SIGGRAPH 2000, New Orleans, Louisiana, July 23--28, 2000}}
\fancyhead[RO]{\fontsize{8}{8}\textsf{Computer Graphics Proceedings, Annual Conference Series, 2000}}

\newlength{\headrulelength}
\setlength{\headrulelength}{\textwidth}
\newlength{\headrulegap}
\setlength{\headrulegap}{1.3in}
\addtolength{\headrulelength}{-\headrulegap}
\def\headrule{\hspace{\headrulegap}\rule[2ex]{\headrulelength}{\headrulewidth}\gdef\headrule{\hrule}}

\fancypagestyle{empty}{
  \fancyfoot{}
  \fancyhead{}
  \fancyhead[RO,LE]{\fontsize{8}{8}\textsf{Computer Graphics Proceedings, Annual Conference Series, 2000}}
  \fancyfoot[RO,LE]{\thepage}
  \fancyhead[LO,RE]{\includegraphics[height=.5in]{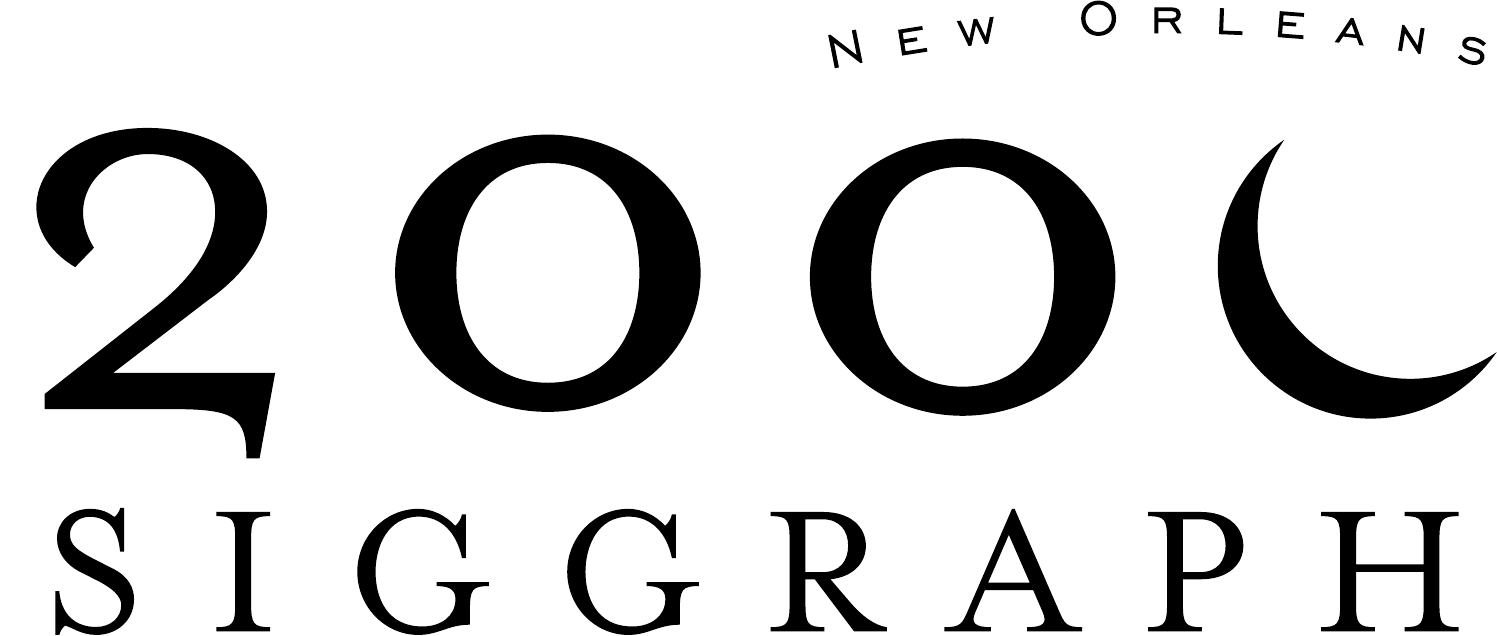}\vspace*{-.45in}}
}

\thispagestyle{empty}\pagestyle{fancy}


\renewcommand{\thefootnote}{}%
\footnotetext[0]{
    \par\noindent 
    College of Computing, Georgia Institute of Technology, Atlanta, GA
    30332.  gyngve@acm.org, job@acm.org, jkh@cc.gatech.edu.\\
    \parbox[t][1.5in][b]{\columnwidth}{\fontsize{8}{8}\textsf{%
      \textbf{SIGGRAPH 2000, New Orleans, LA USA}\\[.15in]
      \textbf{\textit{\color{red} This is an author's preprint.}}\\[.15in]
      \href{https://doi.org/10.1145/344779.344801}{https://doi.org/10.1145/344779.344801}
    }}
}
\renewcommand{\thefootnote}{\arabic{footnote}}
\preprinttext{}  


\section{Introduction}


Explosions are among the most dramatic phenomena in nature.  A sudden
burst of energy from a mechanical, chemical, or nuclear source causes
a pressure wave to propagate outward through the air.  The blast wave
``shocks up,'' creating a nearly discontinuous jump in pressure,
density, and temperature along the wave front.  The wave is
substantially denser than the surrounding fluid, allowing it to travel
supersonically and to cause a noticeable refraction of light.  The air
at the shock front compresses, turning mechanical energy into heat.
The waves reflect, diffract, and merge, allowing them to exhibit a
wide range of behavior.

An explosion causes a variety of visual effects in addition to the
light refraction by the blast wave.  An initial chemical or nuclear
reaction often causes a blinding flash of light.  Dust clouds are
created as the blast wave races across the ground, and massive objects
are moved, deformed, or fractured.  Hot gases and smoke form a rising
fireball that can trigger further combustion or other explosions and
scorch surrounding objects.

\begin{figure}[!t]{\sloppy
  \centerline{\includegraphics[width=\columnwidth]{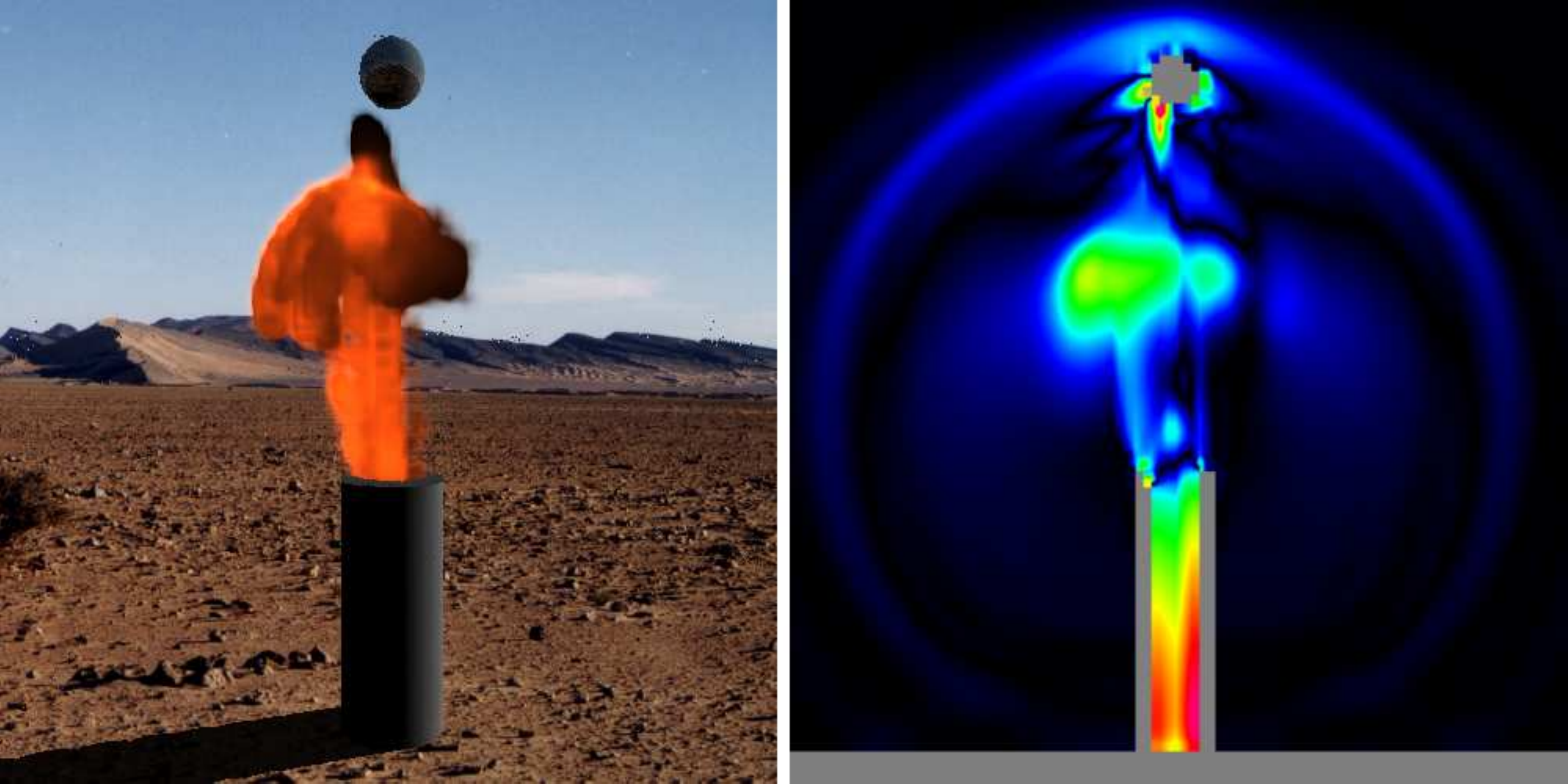}}
  \vspace*{-0.10in} 
  \caption{
    \setlength{\baselineskip}{\captbaselineskip}
    An image of a projectile propelled from a chamber by an explosion.
    On the right is a cross-section of the three-dimensional fluid volume
    using a colormap where hotter colors indicate higher densities.
  }\label{fig:chimney} \vspace*{-0.10in} 
}\end{figure}

We present a physically based model of an explosion and show how it
can be used to simulate many of these effects.  We model the explosion
post-detonation as compressible, viscous flow and solve the flow
equations with an integration method that handles the extreme shocks
and supersonic velocities inherent in explosions.  We cannot capture
many of the visual effects of an explosion in a complex setting if we
rely only on an analytical model of the blast wave; a fluid dynamics
model of the air is necessary to capture these effects.  The system
includes a two-way coupling between dynamic objects and fluid that
allows the explosions to move objects.  Figure~\ref{fig:chimney}
illustrates this phenomenon with a projectile propelled from a
chamber.  We also use the pressure wave generated by the explosion to
fracture and deform objects.  The user can simulate arbitrarily
complex scenarios by positioning polygonal meshes to represent
explosions and objects.  The user controls the scale and visual
qualities of the explosion with a few physically motivated parameters.

\begin{figure*}[!t]{\sloppy
  \centerline{\includegraphics[width=\textwidth]{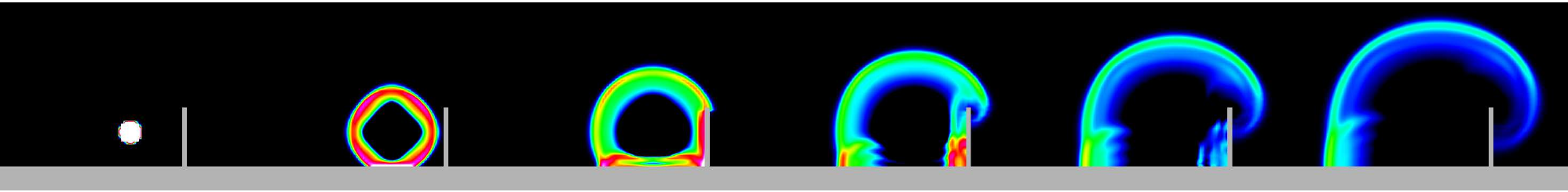}}
  \vspace*{-0.10in} 
  \caption{
    \setlength{\baselineskip}{\captbaselineskip}
    Shown here is a cross-section of pressures for a three-dimensional
    explosion near an immovable wall.  The timestep between frames is
    10\,ms.  Many of the behaviors of the blast wave can be seen,
    specifically the Mach stem formed from the blast wave merging with
    the wave reflected off the ground plane, and the diffracted wave
    formed when the blast wave crests over the wall.
  }\label{fig:wall2d}\vspace*{-0.10in} 
}\end{figure*}

Our fluid model of an explosion simulates many phenomena of blast
waves that existing graphics techniques do not capture.
Figure~\ref{fig:wall2d} shows a cross-section of pressures for a
three-dimensional explosion near a wall.  The initial disturbance in
the first image interacts with the surrounding fluid and causes a
pressure wave to propagate through the medium.  In the second image,
the blast wave has ``shocked up,'' as is evident by the large
differences in pressure across the shock front.  The blast wave
reflects off the wall and the ground in the third image.  In the
fourth, the wave that reflected off the ground merges with
the initial blast wave to form a Mach stem, which has pressure values
twice that of the initial wave.  In the final two images, the blast
wave crests over the wall and forms a weaker diffracted wave.

In the entertainment industry, explosions are currently created at
full scale in the real world, in miniature, or using heuristic
graphics
techniques\cite{Street:1997:volcano,Martin:1998:godzilla,Vaz:1998:armageddon}.
Each of these methods has significant disadvantages, and we believe
that in many scenarios, simulation may provide an easier solution.
When explosions are generated and filmed at full scale in the real
world, they often must be faked to appear dangerous and destructive by
using multiple charges and chemicals with a low flashpoint.  Because
of the cost and danger of exploding full-size objects, many explosions
are created using miniatures.  With miniatures, the greatest challenge
is often scaling the objects and the physics to create a realistic
effect.  Current graphics techniques for creating explosions are based
on heuristics, analytical functions, or recorded data, and although
they produce nice effects for spherical blast waves, they are not
adequate for the complex effects required for many of the scenarios
used in the entertainment industry.

Physically based simulations of explosions offer several potential
advantages over these three techniques.  In contrast to real physical
explosions, simulations can be used in an iterative fashion, allowing
the director many chances to modify or shape the effect.  The
rendering of the explosion is to a large extent decoupled from the
simulation, allowing the visual characteristics of the dust clouds or
fireball to be determined as a post-process.  Unlike heuristic or
analytical graphical methods, physically based simulations allow the
computation of arbitrarily complex scenes with multiple interacting
explosions and objects.

The next section discusses relevant previous work in explosions,
fluids, flame, and fracture.  The following section introduces the
explosion model in the context of computational fluid dynamics.  The
next two sections discuss coupling between the fluid and solids and
other secondary effects such as refraction and fireballs.  We close
with a discussion of our results.
 

\section{Previous Work}

Explosions used in the entertainment industry tend to be visually
rich.  Because of the inherent computational complexity of these
explosions, researchers largely neglected this field after the
publication of particle simulation
techniques\cite{Reeves:1983:PS,Sims:1990:PAA}.  Procedural methods can
generate fiery, billowy clouds that could be used as
explosions\cite{Ebert:1994:TM}.

Recently two papers specifically addressed explosions.  Mazarak and
colleagues simulate the damage done by an explosion to voxelized
objects\cite{Mazarak:1999:AEO}.  They model the explosion as an ideal
spherical blast wave with a pressure profile curve approximated by an
analytic function based on the modified Friedlander equation and
scaled according to empirical laws\cite{Baker:1973:EIA}.  The
spherical blast wave expands independent of existing obstacles, and
forces are applied to objects in the direction of the blast radius.
Objects are modeled as connected voxels and based on various
heuristics, these radial forces may cause the voxels to disconnect.

Neff and Fiume use data from empirical blast curves to model an
explosion\cite{Neff:1999:AVM}.  The blast curves relate the pressure
and velocity of the blast wave to time and are scalable.  Unlike
Mazarak and colleagues, they use a curve representing the reflection
coefficient to apply forces to objects based on the angle of incidence
of the blast wave.  They assume quasi-static loading conditions where
the blast wave encloses the entire object and effects due to reflected
waves are ignored.  They also model explosion-induced fracture in
planar surfaces using a procedural pattern generator.

An alternative to these analytic and empirical models is a
computational fluid dynamic simulation of the blast wave and the
surrounding air.  Foster and Metaxas presented a solution for
incompressible, viscous flow and used it to animate
liquids\cite{Foster:1996:RAO} and hot, turbulent
gas\cite{Foster:1997:MTM}.  They modeled fluid as a three-dimensional
voxel volume with appropriate boundary conditions.  The fluid obeys
the Navier-Stokes equations; gas also follows an equation that
represents thermal buoyancy.  Using an explicit scheme, they update
velocities and temperatures every timestep via Euler integration and
readjust the values to guarantee conservation of mass.  The fluid is
rendered by tracing massless particles along the interpolated flow
field.  Their work with liquids included dynamic objects that were
moved by the fluid, although they assumed that the objects were small
enough not to influence the fluid.  Recently Stam addressed the
computational cost of guaranteeing stability by introducing extra
damping to afford larger timesteps and using an implicit method to
solve a sparse system of equations\cite{Stam:1999:SF}.  Stam's method
is inappropriate for shocks and explosions because his integration
scheme achieves stability by encouraging the fluid to dissipate.

In the dramatic effects produced by the entertainment industry, a
fireball is often the most salient visible characteristic of an
explosion.  Stam and Fiume modeled flame and the corresponding fluid
flow and rendered the results using a sophisticated global
illumination method\cite{Stam:1995:DFA}.  The gases behaved according
to advection-diffusion equations; Stam and Fiume solve these equations
efficiently by reformulating the problem from a grid to ``warped
blobs.'' Illumination from gas is affected by emission and anisotropic
scattering and absorption.  They only consider continuous emissions
from blackbody radiation and ignore line emissions from electron
excitation.  They develop a heuristic for smoke emission due to the
lack of a scientific analytic model.


Compressible flow has been studied for years in the computational
fluid dynamics
community\cite{Anderson:1990:MCF,Baker:1973:EIA,Madder:1979:NMD}.  We
have built on this work by taking the governing equations and the
donor-acceptor method of integration from this literature. However,
the reasons for simulating explosions, combustion, detonation, and
supersonic flow in engineering differ significantly from those in
computer graphics.  Engineering problems often require focusing on one
element such as the boundary layer and simulating the other elements
only to the extent that they affect the phenomenon under study.  For
example, engineering simulations are often two-dimensional and assume
symmetry in the third dimension.  Because they are focused on a
specific event, their simulations may run for only a few microseconds.
In computer graphics, on the other hand, we need to produce a visually
appealing view of the behavior throughout the explosion. As a result,
we need a more complete model with less quantitative accuracy.


\section{Explosion Modeling}

An explosion is a pressure wave caused by some initial disturbance,
such as a detonation.  In the results presented here, we assume that
the detonation has occurred and that its properties are defined in the
initial conditions of the simulation.  This assumption is reasonable
for most chemical explosions because the detonation is complete within
microseconds.  We animate explosions by modeling the pressure wave and
the surrounding air as a fluid discretized over a three-dimensional
rectilinear grid.  The following two subsections describe the
governing equations for fluid dynamics and the computational
techniques used to solve them.  The remaining two subsections describe
the parameters available to the user for controlling the appearance of
the explosion via the boundary conditions and initial conditions.


\subsection{Fluid Dynamics}

In nearly all engineering problems, including the analysis of
explosions, fluids are modeled as a continuum.  They are represented
as a set of equations in terms of density
$\rho$\,($\RM{kg}/\RM{m}^3$), pressure 
$P$\,($\mathrm{N}/\mathrm{m}^2$), velocity
${\BM{v}}$\,($\mathrm{m}/\mathrm{s}$), temperature
$T$\,($\mathrm{K}$), the internal energy per unit mass
$N$\,($\mathrm{J}/\mathrm{kg}$), and the total energy per unit mass
$E=N+\frac{1}{2}\BM{v}^2$\,($\mathrm{J}/\mathrm{kg}$).  The equations
that govern these quantities are defined in an Eulerian fashion,
that is, they apply to a differential volume of space that is
filled with fluid rather than to the fluid itself.  In addition to the
Navier-Stokes equations, which model the conservation of momentum, the
equations for compressible, viscous flow include governing equations
for the conservation of mass and energy and for the fluid's thermodynamic
state\cite{Kuethe:1998:FAB}.

We introduce several simplifying assumptions that make the equations
easier to compute but nevertheless allow us to capture the effects of
compressible, viscous flow.  We discount changes in the vibrational
energies of molecules and assume air to be at chemical equilibrium; we
ignore the effects from dissociation or ionization.  These
assumptions, which are commonly used in the engineering
literature\cite{Anderson:1990:MCF}, allow us to reduce to constants
many of the coefficients that vary with temperature.  The resulting
deviation in the values of the coefficients is negligible at
temperatures below $1000\,\RM{K}$; only minor deviations occur below
$2500\,\RM{K}$.  Our implementation produces aesthetic results with
temperatures above $100000,\RM{K}$, although deviations in constants
could be on the order of a magnitude or two.

The first governing equation of fluid dynamics arises from the conservation 
of mass.  Because fluid mass is conserved, the change of fluid density in a
differential volume must be equal to the net flux across the volume's
boundary, giving
\begin{equation} \label{eq:gov1}
  \frac{\partial \rho}{\partial t} = -\nabla \cdot (\rho {\BM{v}}).
\end{equation}

The second governing equation, commonly known as the Navier-Stokes
equation, concerns the conservation of momentum.  For a Stokes fluid,
where the normal stress is independent of the rate of dilation, the
equation for the $x$ component of the fluid velocity is given by
\begin{equation} \label{eq:gov2}
  \rho \frac{\partial\BM{v}_x}{\partial t} = 
    \rho \BM{f}_x 
  - \nabla P 
  + \frac{\mu}{3}\nabla \cdot \left( \frac{\partial\BM{v}}{\partial x} \right)
  + \mu  \nabla^2 \BM{v}_x
  - \rho(\BM{v} \cdot \nabla) \BM{v}_x 
  ,
\end{equation}
where $\BM{f}$ represents the body forces such as gravity and $\mu$ is
the coefficient of viscosity.  The equations for the $y$ and $z$
components are similar.  The first two terms on the right-hand side of
the equation model accelerations due to body forces and forces arising
from the pressure gradient; the next two terms model accelerations due
to viscous forces.  The last term is not a force-related term; rather
it is a convective term that models the transport of momentum as the
fluid flows.  This distinction between time derivative (force) terms
and convective terms will be important for the integration scheme.

The final governing equation enforces the conservation of energy in
the system.  The First Law of Thermodynamics dictates that the change
in energy is equal to the amount of heat added and the work done to
the system.  Accounting for the amount of work done from pressure and
viscosity and the heat transferred from thermal conductivity yields
\begin{equation}\label{eq:gov3}
  \rho \frac{\partial N}{\partial t} =
  k \nabla^2 T - P \nabla \cdot \BM{v} + \Phi
  - \rho(\BM{v} \cdot \nabla)N,
\end{equation}
where $k$ is the thermal conductivity constant and $\Phi$ is the
viscous dissipation given by
\begin{equation}
  \Phi = -\frac{2\mu}{3} (\nabla \cdot \BM{v})^2 +
    \frac{\mu}{2}
    \sum_{i,j \in \{ x,y,z \}}
      \left(\frac{\partial\BM{v}_i}{\partial j} + \frac{\partial\BM{v}_j}{\partial i}\right)^2 .
\end{equation} 
\par\noindent 
As with equation~(\ref{eq:gov2}), the last term of
equation~(\ref{eq:gov3}) is a convective term and models the
transport of energy as the fluid flows.

In addition to the three governing equations, we need 
equations of state that determine the relationship between energy,
temperature, density, and pressure.  They are
\begin{eqnarray}\label{eq:state1}
  N = c_V T,\;\;P = \rho R T,
\end{eqnarray}
where the coefficient $c_V$ is the specific heat at constant volume and $R$
is the gas constant of air.


\begin{figure}[!t]{\sloppy
  \centerline{\includegraphics[width= 0.75 \columnwidth]{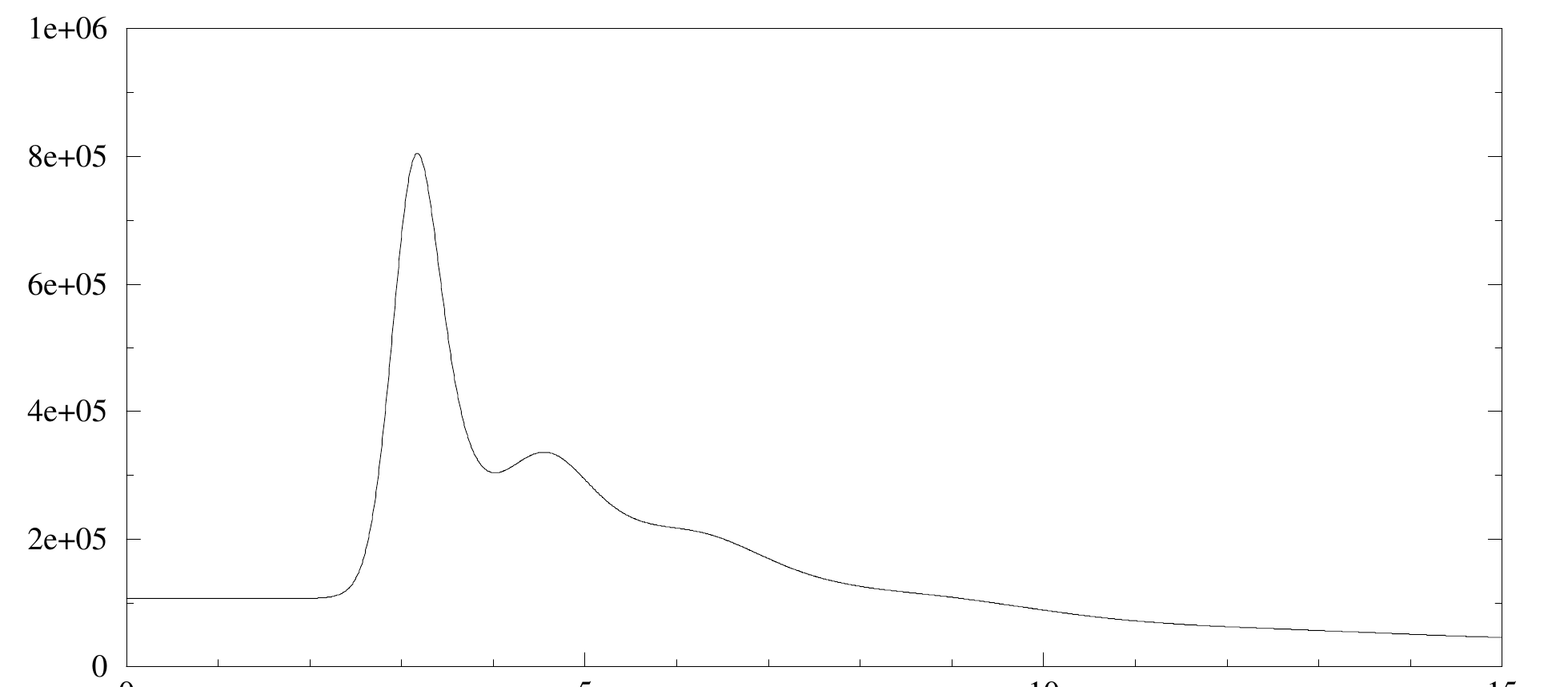}}
  \vspace*{-0.10in} 
  \caption{
    \setlength{\baselineskip}{\captbaselineskip}
    Pressure profile ($\mathrm{N}/\mathrm{m}^2$) over time ($\mathrm{ms}$) near an
    explosion.
  }\label{fig:shock} \vspace*{-0.10in} 
}\end{figure}

\begin{figure}[!b]
  \centerline{\includegraphics[width= 0.75 \columnwidth]{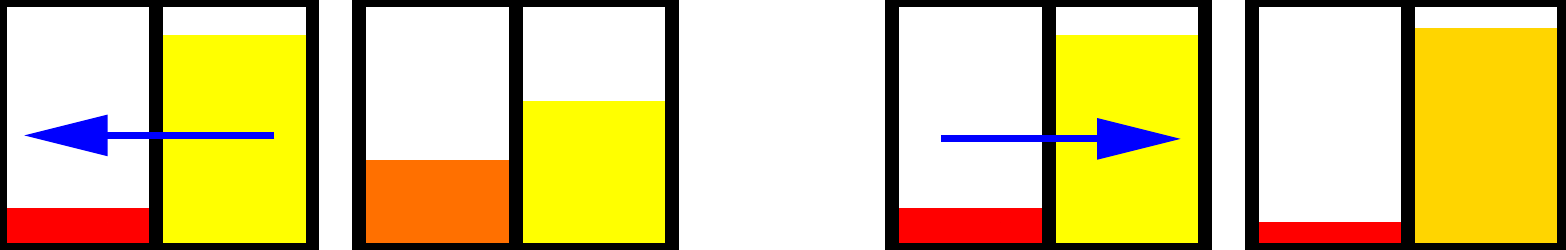}}
  \vspace*{-0.10in} 
  \caption{
    \setlength{\baselineskip}{\captbaselineskip}
    This figure illustrates the donor-acceptor method in which the
    amount of mass transferred is proportional to the mass of the
    donor.  The voxels on the left show the transfer of mass and
    energy according to the flow, indicated by the blue arrow.  The
    two voxels on the right represent the scenario with reversed flow
    of the same magnitude.  Density is represented as height, and unit
    energy is represented as color. Corresponding amounts of energy
    are sent with the mass.  
  }\label{fig:donor} \vspace*{-0.10in}
\end{figure}

\subsection{Discretization and Numerical Integration}
\label{sec:fluiddyn}
The equations in the previous section describe the behavior of a
fluid in a continuous fashion.  However, implementing them in a form
suitable for numerical computation requires that the space filled
by the fluid be discretized in some manner and that a stable method for
integrating the governing equations forward in time be devised.

Finite differences are used to discretize the space into a regular
lattice of cubical cells.  These finite voxels take the place of the
differential volumes used to define the continuous equations, and the
governing equations now hold for each voxel.  Fluid properties such as
pressure and velocity are associated with each voxel and these
properties are assumed to be constant across the voxel.  The spatial
derivatives used in the governing equations are approximated on the
lattice using central differences.  For example, the $x$ component of
the pressure gradient, $\nabla P$, at voxel $[i,j,k]$ is given by
\begin{equation}
  \frac{\partial P}{\partial x} \approx
  \frac{P_{[i+1,j,k]}-P_{[i-1,j,k]}}{2 h},
\end{equation}
where subscripts in square brackets index voxel locations and $h$ is
the voxel width.

After the governing equations have been expressed in terms of discrete
variables using finite differences, they may be used as the update
rules for an explicit integration scheme.  However, rapid pressure
changes created by steep pressure gradients moving at supersonic speeds
would cause such a scheme to diverge rapidly.  (See
Figure~\ref{fig:shock}.)  To deal with this problem, we improved the
basic integration technique using two modifications described in the
fluid dynamics literature\cite{Baker:1973:EIA,Madder:1979:NMD}.  The
first modification involves updating equations~(\ref{eq:gov2})
and~(\ref{eq:gov3}) in two steps, first using only the temporal
portion of the derivatives and second using the convective
derivatives.  The second modification is called the
\textit{donor-acceptor method} and is described in detail below.  It
addresses problems that arise when mass, momentum, and energy are
convected across steep pressure gradients.

The modified update scheme operates by applying the following
algorithm to each voxel at every timestep:
\begin{enumerate}
  \setlength{\itemsep}{0.01in}
  \item Approximate the fluid acceleration at the current time,
  $\BM{\widetilde{a}}_t =
  \left({\partial{\BM{v}}}/{\partial{t}}\right)_t$, using the
  non-convective (first four) terms of equation~(\ref{eq:gov2}).
  \item Compute the tentative velocity at the end of the timestep,
  $\BM{\widetilde{v}}_{t+\Delta t} = \BM{v}_t + \Delta t
  \BM{\widetilde{a}}_t$, and the approximate average velocity during the
  timestep $\BM{\bar{v}}_t = (\BM{\widetilde{v}}_{t+\Delta t} + \BM{v}_t)/2$.
  \item Approximate change in internal energy, $N$, using the
  non-convective terms of equation~(\ref{eq:gov3}) and substituting
  $\BM{\bar{v}}_t$ for the fluid velocity.
  \item  Using $\BM{\bar{v}}_t$ for the fluid velocity, compute the new density, $\rho_{t+\Delta t}$ with equation~(\ref{eq:gov1}).
  \item Calculate the complete $\BM{{v}}_{t+\Delta t}$ and
  $N_{t+\Delta t}$ with equations~(\ref{eq:gov2}) and~(\ref{eq:gov3})
  using the convective terms and the new value of $\rho$.
  \item Use state equations~(\ref{eq:state1}) to
  update secondary quantities such as temperature.
\end{enumerate}

Although this update scheme is more stable than a simple Euler
integration, sharp gradients in fluid density may still allow small
flows from nearly empty voxels to generate negative fluid densities
and cause inappropriately large changes to both velocity and internal
energy.  To prevent these problems, we use a donor-acceptor method
when computing $-\nabla \rho\BM{v}$ of the convective terms in steps~4
and~5 above.

The donor-acceptor method transfers mass proportional to the mass of
the voxel in the upstream direction, or donor voxel.  Suppose we have
voxel $i$ and one of its six neighbors $j$ in direction $\BM{d}$ from
$i$.  Let $v_{ij} = \frac{1}{2}(\BM{v}_i+\BM{v}_j)\cdot \BM{d}$. If
$v_{ij}>0$, then flow is going from $i$ to $j$, $i$ is the donor, and
$\rho_i$ is used in equation~(\ref{eq:gov1}) to compute the change in
voxel $i$'s density.  Likewise, if $v_{ij}<0$, then $j$ is the donor,
and the density of $j$ is used when computing the change in voxel
$i$'s density.  These calculations are repeated for the six neighbors
of $i$ to obtain the new density for $i$, $\rho_{t+\Delta t}$.  The
velocity and energy convection are then scaled by
${\rho_t}/{\rho_{t+\Delta t}}$ to conserve momentum and energy.
Figure~\ref{fig:donor} illustrates the donor-acceptor method.  The
left and right diagrams show flows in opposite directions of the same
magnitude.  The sent mass is proportional to the mass of the donor
and carries with it corresponding amounts of energy.


\begin{figure}[!t]{\sloppy
  \centerline{\includegraphics[width=\columnwidth]{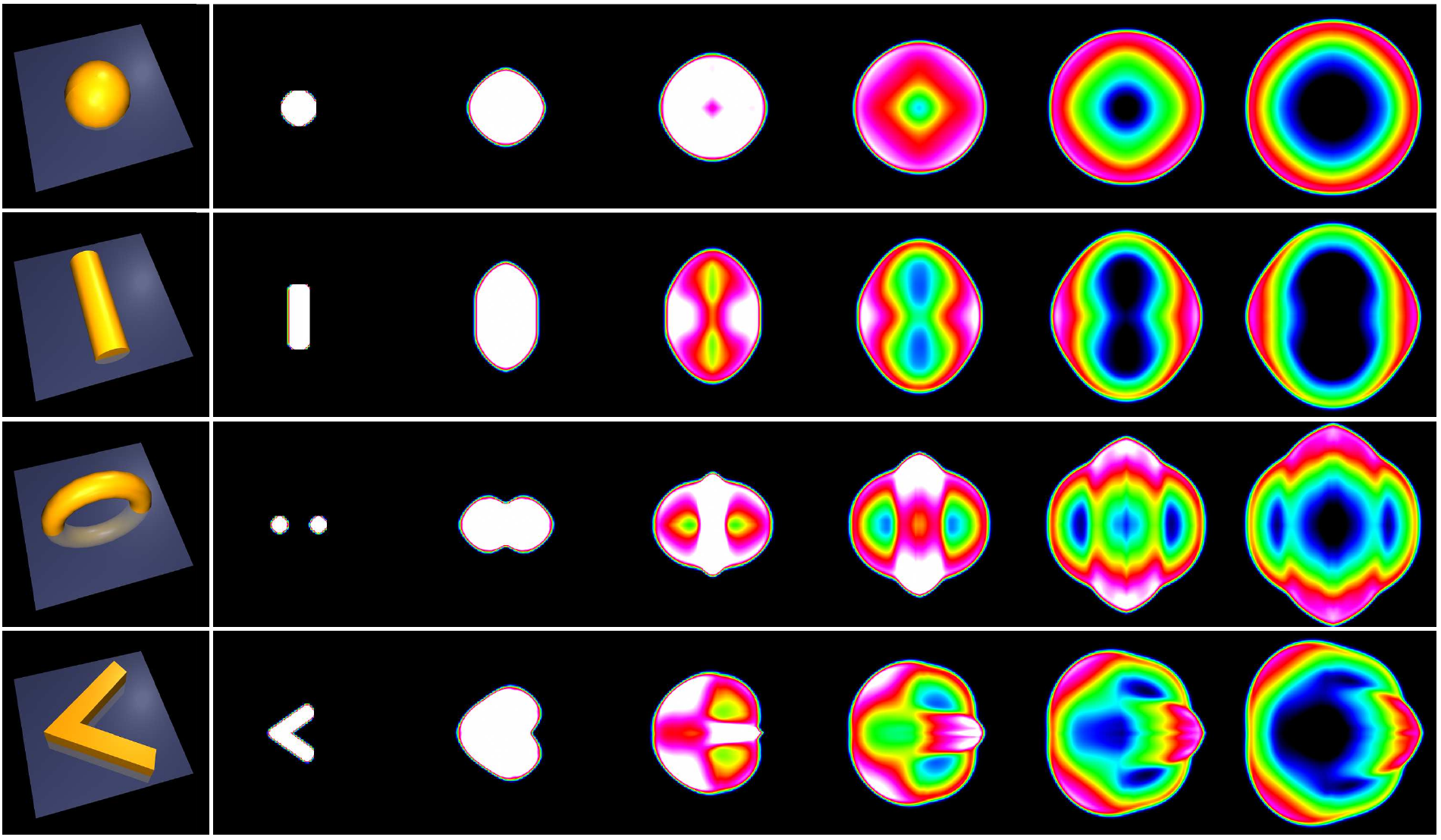}}
  \vspace*{-0.10in} 
  \caption{ 
    \setlength{\baselineskip}{\captbaselineskip}
    This figure shows cross-sections of pressure for three-dimensional
    explosions of equal-volume charges in the shape of a sphere,
    cylinder, torus, and wedge.  The timestep between frames is 5\,ms.
  }\label{fig:shaped}\vspace*{-0.10in}
}\end{figure}

\subsection{Boundary Conditions}
\label{sec:bound}
The system has several types of boundary conditions that allow the
fluid to exhibit a wide range of behaviors.  Free boundaries allow
blast waves to travel beyond the voxel volume as if the voxel volume
were arbitrarily large.  This type of boundary allows us to model
slow, long-term aspects of explosions, such as fireballs and dust
clouds.  Hard boundaries force fluid velocity normal to them to be
zero while leaving all other fluid attributes unchanged.  We treat
these boundaries as smooth surfaces, so tangential flow is unaffected.
We implement a third boundary condition to achieve faster execution.
If a voxel and its neighbors have pressure differences less than a
threshold, the voxel is treated as a free boundary and is never
evaluated.  This optimization prunes out the majority of the volume
while the blast wave is expanding.


\subsection{Initial Conditions}

The user specifies the pressure and temperature of the air, and the
initial values of other variables are determined from the state
equations~(\ref{eq:state1}).  The detonation is initialized by
specifying a region of the volume with higher temperature or pressure.
For example, a chemical explosion might have a temperature of
$2900\,\RM{K}$ and a pressure of $1000\,\RM{atm}$ with the surrounding
air at $290\,\RM{K}$ and $1\,\RM{atm}$.  The creation of the explosion
may be time-delayed or may be triggered when the fluid around the
charge reaches a threshold temperature.

The detonation may have an arbitrary geometry represented by a
manifold polygonal mesh.  The mesh is voxelized to initialize the
appropriate voxels in the fluid simulation.  By controlling the
geometry, the user can produce a variety of effects that could not be
achieved with a spherical model.  In blast theory, planar, cylindrical, and
spherical blast waves can be modeled by analytic functions\cite{Baker:1973:EIA}; however
nonstandard shapes can create surprising and interesting effects.
Figure~\ref{fig:shaped} shows cross-sections of pressure for
three-dimensional explosions from equal-volume charges in the shape of
a sphere, cylinder, torus, and wedge.  The inner blast wave of the
torus merges to create a strong vertical blast wave.  The wedge
concentrates its force directly to the right, while leaving the
surrounding area relatively untouched.


\section{Interaction with Solids}

People use explosions to impart forces on objects for both
constructive and destructive purposes.  The movements of these objects
and the resulting displacement of air create many of the compelling
visual effects of an explosion.  In this section, we present methods
to implement a two-way coupling between the fluid and solids.  The
coupling from fluid to solid allows us to model phenomena such as a
projectile being propelled by an explosion.  The coupling from solid
to fluid can be used to model a piston compressing or the shock wave
formed as a projectile moves through the air supersonically.  We
also extend previously published techniques for fracture to allow the
pressure wave to shatter objects.

To allow the two-way coupling, objects have two representations: a
polygonal mesh that is used to apply forces to the object from the
fluid, and a volume representation in voxels that is used to displace
fluid based on the motion of the object.  We incorporate the coupling
into the fluid dynamics code in the following way:
\begin{enumerate}
  \setlength{\itemsep}{0.01in}
  \item Apply forces on the objects from the fluid and compute the
  rigid body motion of the objects.
  \item If the object has moved more than a fraction of a voxel,
  recompute the voxelization of the object.
  \item Displace fluid based on object movement.
  \item Update the fluid using the techniques described in Section~\ref{sec:fluiddyn}.
\end{enumerate}
We explain the first three of these items in greater detail in the
following subsections.


\subsection{Coupling from Fluid to Solid}

An object embedded in a fluid experiences two separate sets of forces
on its surface, those arising from hydrostatic pressure and those
arising from dynamic forces due to fluid momentum.  The forces due to
hydrostatic pressure act normal to the surface and are generated by
the incoherent motions of the fluid molecules against the surface.
The dynamic forces are generated by the coherent motion of the
continuous fluid and can be divided into a force normal to the surface
of the object and a tangential shearing force.  We neglect the
tangential shearing force because in the context of explosions, it is
negligible in comparison to the force due to hydrostatic pressure.  We
assume that the object is in equilibrium under ambient air pressure
and the hydrostatic forces are computed using the overpressure
$\bar{P}$, which is the difference between the hydrostatic pressure
$P$ and ambient pressure.

The magnitude of the normal force per unit area on the surface is
given by the dynamic overpressure:
\begin{equation}
  \bar{P}_\mathit{dyn}=\bar{P}+\frac{1}{2}\rho
  \left(\BM{v}_\mathit{rel} \cdot \BM{\hat{n}} \right)^2,
\end{equation}
where $\BM{v}_\mathit{rel}$ is the velocity of the fluid relative to
the surface, and $\BM{\hat{n}}$ is the outward surface normal.

We assume that the triangles composing each object are small enough
that the force is constant over each triangle.  The force on a
triangle with area $A$ is then
\begin{equation}
  \BM{f}=-\BM{\hat{n}} A \bar{P}_\mathit{dyn},
\end{equation}
where the fluid properties are measured at the centroid of the
triangle.  The forces are computed for all triangles of an object, and
the translational and angular velocities of the object are updated
accordingly.

\begin{figure}[!t]
  \centerline{\includegraphics[width=\columnwidth]{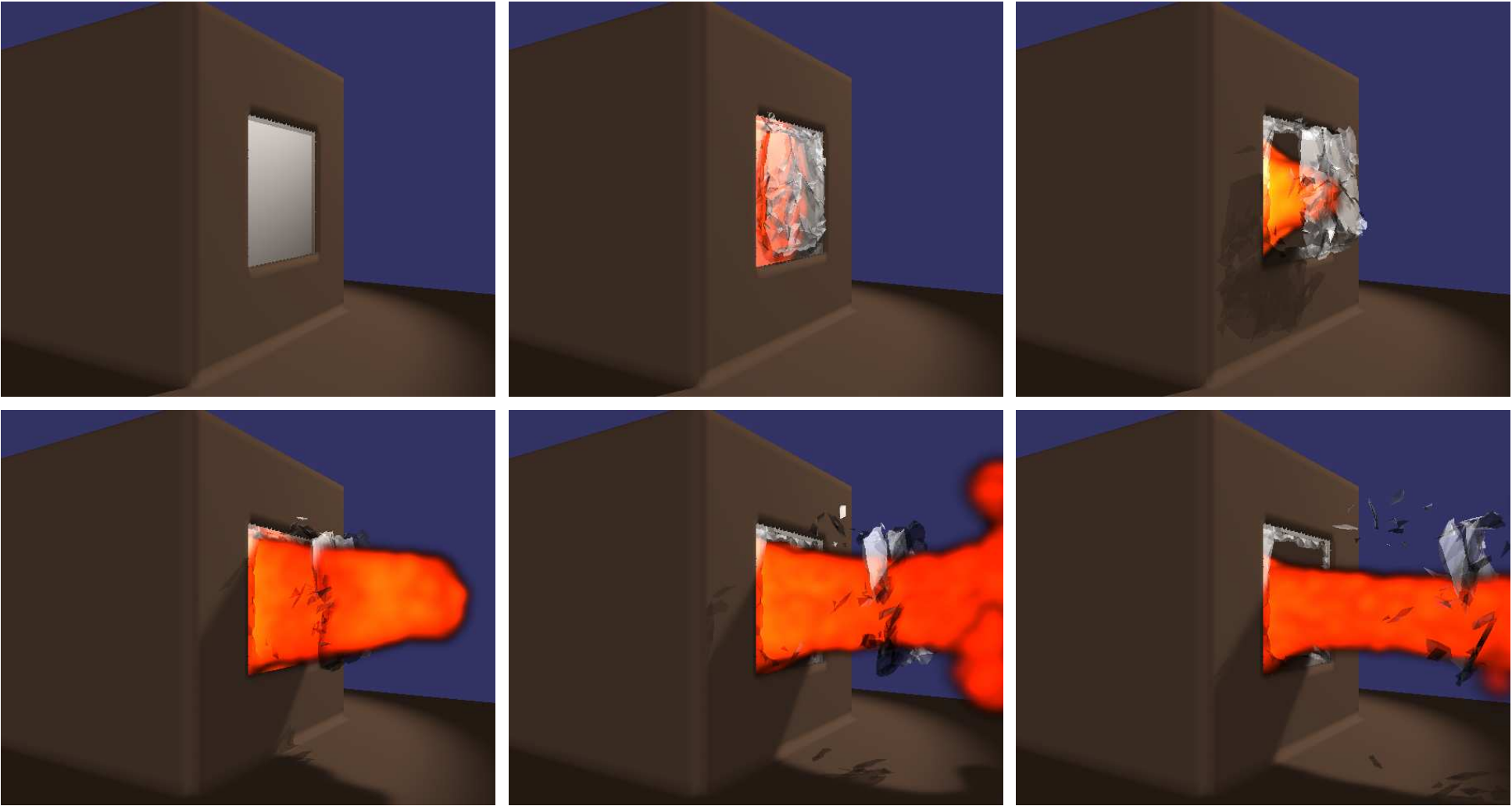}}
  \vspace*{-0.10in} 
  \caption{
    \setlength{\baselineskip}{\captbaselineskip}
    A glass window is shattered by a blast wave.  The blast
    wave pressure is approximately $3\,\RM{atm}$ when it reaches the
    window.  The images show the scene at 0\,ms, 13\,ms, 40\,ms, 67\,ms, 107\,ms, and 160\,ms. 
  }\label{fig:window}
  \vspace*{-0.10in}
\end{figure}

\begin{figure}[!b]
  \centerline{\includegraphics[width=\columnwidth]{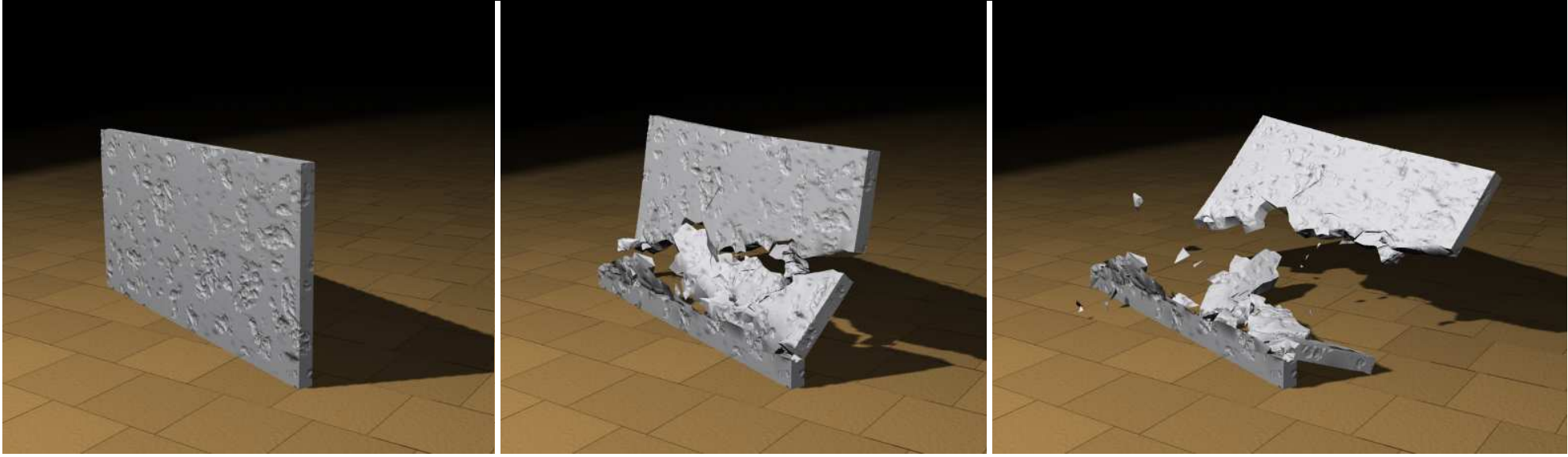}}
  \vspace*{-0.10in} 
  \caption{
    \setlength{\baselineskip}{\captbaselineskip}
    A wall is knocked over by a blast wave from an explosion 3.5\,m
    away.  The images are spaced 250\,ms apart.
  }\label{fig:wallBreak} \vspace*{-0.10in}
\end{figure}

In addition to acting on rigid objects, the forces can also be applied
to flexible objects that deform and fracture\cite{OBrien:1999:GMA}.
The explosion simulation results in pressures, velocities, and
densities for each voxel in the discretization of the fluid.  The
fracture simulation uses this information to compute the forces that
should be applied to a finite element model of the objects in the
scene.  The force computation is similar to that for rigid objects.
This method was used to simulate the breaking window shown in
Figure~\ref{fig:window} and the breaking wall shown in
Figure~\ref{fig:wallBreak}.  This coupling is one-way in that the
fluid applies forces to the finite-element model, but the fluid is not
moved by the fragments that pass through it.


\subsection{Coupling from Solid to Fluid}

To allow the solid to displace fluid, the triangular mesh representing
the object is converted to voxels\cite{Nooruddi:1999:SRP}, which are
then used to define the hard boundaries in the fluid volume
dynamically.  The objects move smoothly through the fluid, but because
of the discrete nature of the voxelization, large changes in the
amount of fluid displaced may occur on each timestep.  To address this
problem, the fluid displaced or the void created by the movement of
the objects is handled over a period of time rather than
instantaneously.

The voxelization returns a value between zero and one representing the
proportion of the voxel that is not interior to any object.  This
value is independent of geometric considerations about the exact
shape of the occupied volume.  If any dimension of an object is
smaller than the size of a voxel, the appropriate voxels will have
partial volumes, but because there are no fully occupied voxels, the
fluid will appear to move through the object.  Nonzero partial
volumes below a certain threshold are set to zero to increase stability.
The implementation of partial volumes requires slight modifications to
the donor-acceptor method to conserve mass, momentum, and energy
because two adjacent voxels could have different volumes.

When any part of an object moves more than a fraction of a voxel, the
object is revoxelized, and the hard boundaries of the fluid are
updated.  When this process occurs, the partial volume in a voxel
might change, resulting in fluid flow.  We allow this flow to occur
smoothly by sacrificing conservation of mass and energy in the
short-term.  The voxelization determines the partial volumes in an
instantaneous fashion, but the fluid displacement routine maintains
internal partial volumes that change more slowly and are used to
compute the pressure, density, and temperature of the affected voxels.
The internal partial volumes change proportionally to the velocities
of the moving objects, and mass and energy are restored over time.

To compute a smooth change in the internal partial volume from $V_1$
to $V_2$, we model an object moving into a voxel as a piston
compressing or decompressing fluid.  We simplify the computation of
the change in partial volume by assuming that the piston is acting
along one of the axes of the voxelization.  The appropriate axis is
selected based on the largest axial component of the velocity of the
object, $\BM{v}_p$.  The displacement of the piston after $t$ seconds
and the corresponding change in partial volume of a voxel with width
$h$ are
\begin{equation} 
\BM{l}=\BM{v}_p t,\;\;\Delta V=V_2-V_1=h^2\BM{v}_p t. 
\end{equation} 
The displacement occurs linearly over 
\begin{equation} 
t = {\Delta V \over h^2\BM{v_p}} 
\end{equation} 
at a velocity of $\BM{v}_p$.  

Given this model of the change in
internal partial volume, we know that $\rho_1V_1=\rho_2V_2$ because
mass is conserved.  However, the fluid is compressible, so mechanical
energy is not conserved (otherwise $P_1V_1=P_2V_2$).  To obtain the
new pressures and densities of the fluid, we use a thermodynamic
equation relating the work done to the system from changing the volume
(or density),
\begin{equation}
 \frac{P_2}{P_1}=\left(
 \frac{\rho_2}{\rho_1} \right)^\gamma=\left( \frac{T_2}{T_1}
 \right)^{\gamma/(\gamma-1)},
\end{equation}
where $\gamma=1+R/c_V$ ($\gamma$ is about~1.4 for air and is closer
to~1 the more incompressible the fluid)\cite{Anderson:1990:MCF}.
Internal unit energy and total unit energy are then updated by the
state equations.

When the partial volume of a voxel changes from one nonzero value to
another, the resulting pressure changes cause fluid to move to or from
a neighbor based on the governing equations.  However, when the
partial volume of a voxel changes from zero to nonzero or vice versa,
the situation must be handled as a special case by treating the
affected voxel and one of its neighbors as a single larger voxel with
a nonzero partial volume.  The neighbor is selected based on the
largest axial component of the object's velocity, $\BM{v}_p$.  We
calculate the internal partial volume for each of the involved voxels
$A$ and $B$ as $\widetilde{V}_{A_1}$ and $\widetilde{V}_{B_1}$.

When $V_{A_1}$, the original partial volume of $A$, is zero and $V_{A_2}$, the
new partial volume of $A$, is nonzero, we use initial
volumes $\widetilde{V}_{A_1}$ and $\widetilde{V}_{B_1}$ such that
$\widetilde{V}_{A_1}+\widetilde{V}_{B_1}=V_{B_1}$ (the initial volume
is conserved) and
$\widetilde{V}_{A_1}{V}_{B_2}=\widetilde{V}_{B_1}V_{A_2}$ (the voxels
are treated as a single larger voxel).  The change in volume is
\begin{eqnarray}
\Delta{V}_A & = & {V}_{A_2}-\widetilde{V}_{A_1}  =  {V}_{A_2}-{V}_{A_2}\frac{V_{B_1}}{{V}_{A_2}+{V}_{B_2}}, \\
\Delta{V}_B & = & {V}_{B_2}-\widetilde{V}_{B_1}  =  {V}_{B_2}-{V}_{B_2}\frac{V_{B_1}}{{V}_{A_2}+{V}_{B_2}}.
\end{eqnarray} 

When the original partial volume of $A$, $V_{A_1}$, is nonzero and
the new partial volume of $A$, $V_{A_2}$, is zero, we force
$\widetilde{V}_{A_1}$ to be zero and treat $\widetilde{V}_{B_1}$ as a
single larger voxel.  To treat the two voxels as one, we first average
the properties of $A$ and its neighbor $B$, transferring any lost
kinetic energy to internal energy.  The change in volume is then
\begin{equation}
\Delta{V}_B  =  V_{B_2}-\widetilde{V}_{B_1}=V_{B_2}-{\rho_{A}{V}_{A_1}+\rho_{B}{V}_{B_1} \over \rho_B}
\end{equation}
by making sure that $\rho_B\widetilde{V}_{B_1}=\rho_A V_{A_1}+\rho_B
V_{B_1}$ (mass is conserved).


\section{Secondary Effects}

\begin{figure}[!t]
  \centerline{\includegraphics[width=\columnwidth]{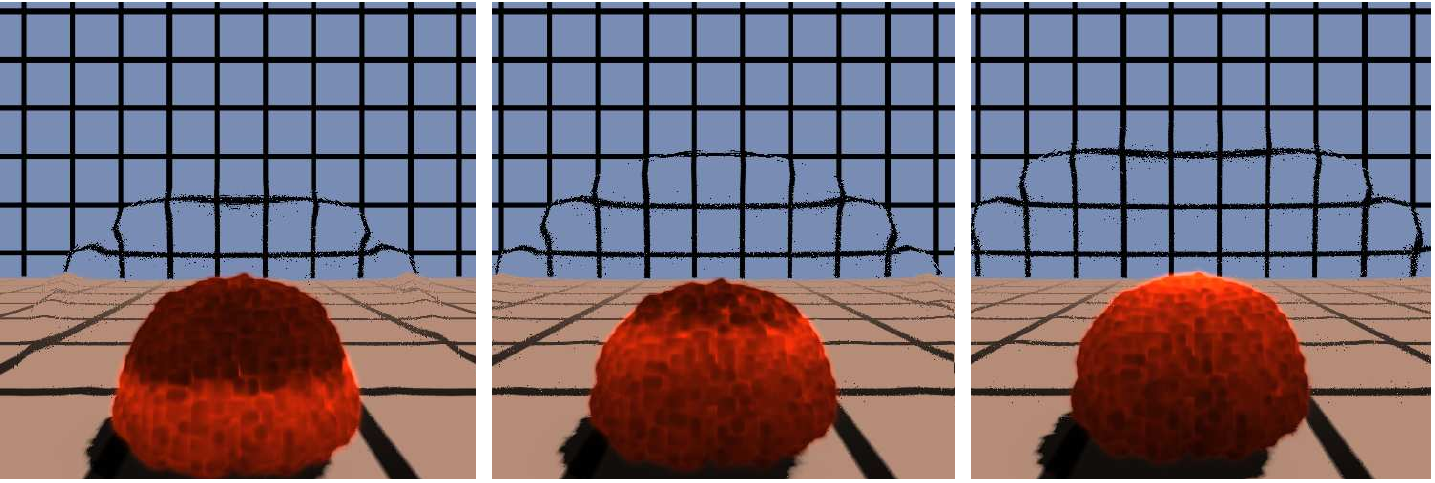}}
  \vspace*{-0.10in} 
  \caption{
    \setlength{\baselineskip}{\captbaselineskip}
    Refraction of light from a blast wave.  Each frame is 10\,ms apart.
    The index of refraction is exaggerated tenfold to enhance the
    effect.  
  }\label{fig:grids}
  \vspace*{-0.10in}
\end{figure}

\begin{figure}[!b]
  \centerline{\includegraphics[width=\columnwidth]{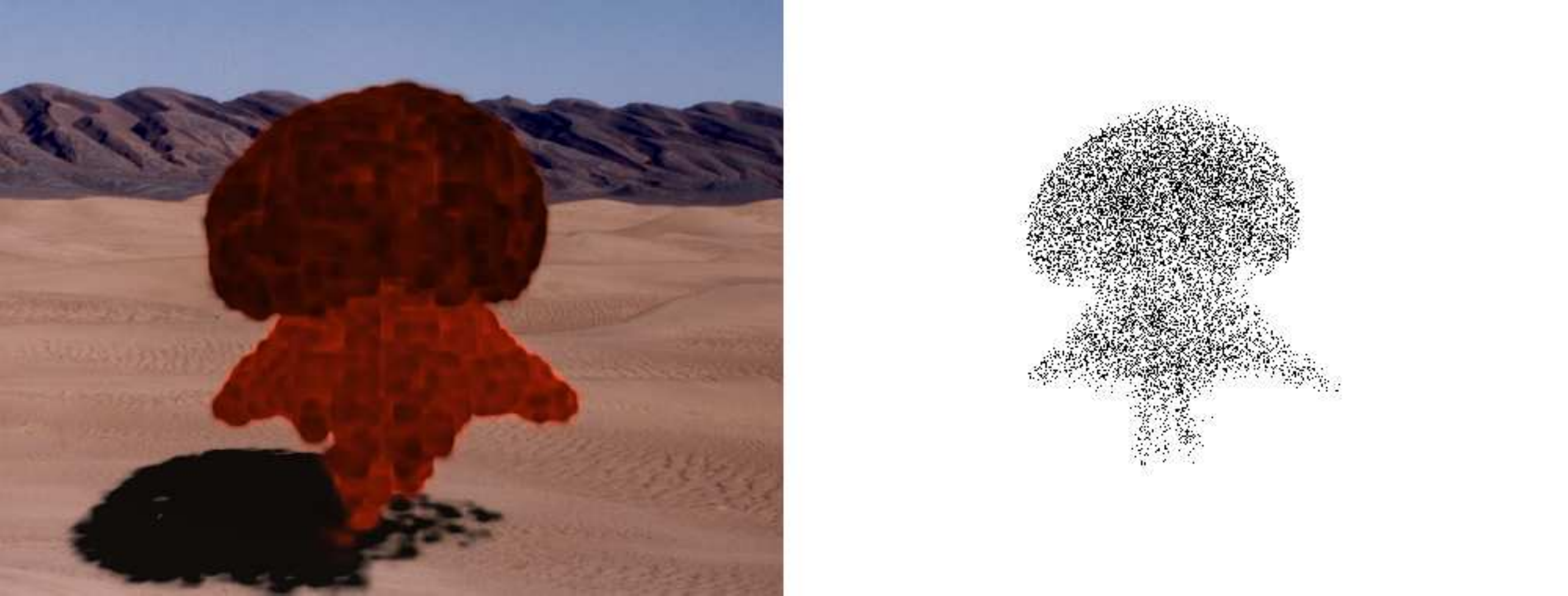}}
  \vspace*{-0.10in} 
  \caption{
    \setlength{\baselineskip}{\captbaselineskip}
    A fireball after one second of simulation time.  Tracer particles
    from the fluid simulation determine the position and coloration of
    the fireball.
  }\label{fig:fireball}
  \vspace*{-0.10in}
\end{figure}

An explosion creates a number of visual secondary effects including
the refraction of light, fireballs, and dust clouds.  These secondary
effects do not significantly affect the simulation, so they can be
generated and edited as a post-process.

One of the most stunning, but often ignored, effects of an explosion
is the bending of light from the blast wave.  Because the blast wave
is substantially denser than the surrounding air, it has a higher
index of refraction, $\eta$.  Light travels at the same velocity
between molecules, but near molecules it is slowed down from
interactions with electrons.  This concept is expressed numerically as
$\eta-1 = k\rho,\;k=2.26\times10^{-4}\;\mathrm{m}^3/\mathrm{kg}$, by
the Dale-Gladestone law\cite{Meyer-Ardent:1984:ICM}.  We capture the
refraction of light by ray tracing through the fluid volume.  (See
Figure~\ref{fig:grids}.)  As the ray is traced through the volume, the
index of refraction is continually updated based on the interpolated
density of the current position.  For simplicity, we compute the
density of each point using a trilinear interpolation of the densities
of the neighboring voxels.  When the index of refraction changes by
more than a threshold, the new direction of the ray is computed via
Snell's law using the density gradient as the surface normal.  The
trilinear interpolation results in minor faceting effects that cause
small errors in the reflected direction.

\begin{figure}[!t]
  \centerline{\includegraphics[width=\columnwidth]{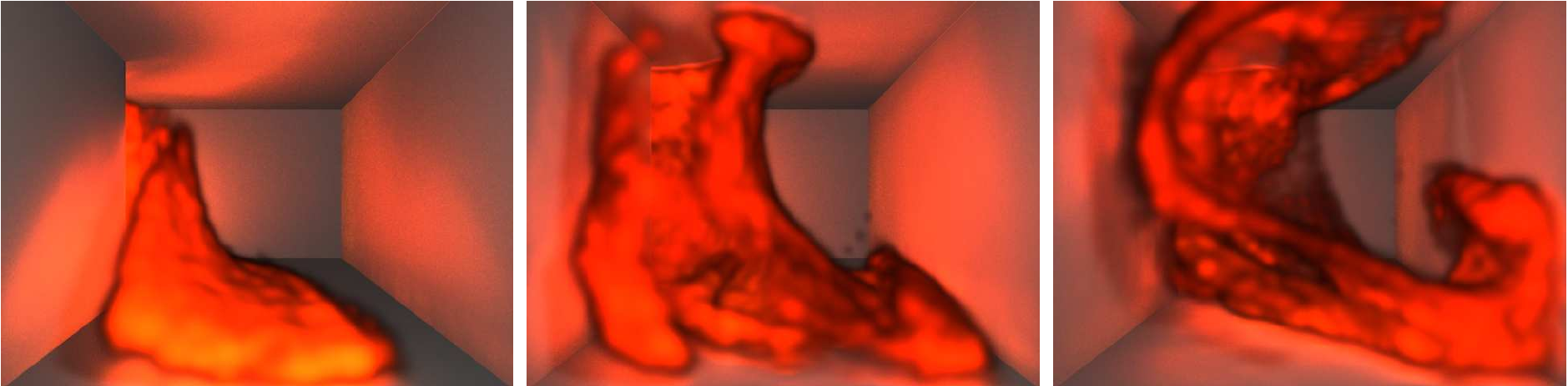}}
  \vspace*{-0.10in} 
  \caption{ 
    \setlength{\baselineskip}{\captbaselineskip}
    A fireball coming around a corner.  The images are spaced 333\,ms apart.
  }\label{fig:corner}
  \vspace*{-0.10in}
\end{figure}

An advantage of using a full volumetric fluid representation for
explosions is that the simulation can be used to model a fireball in
addition to the blast wave.  We assume that the fireball is composed
of detonated material from inside the explosive.  To track this
material, the system initializes the fireball by placing particles
inside the shape specified by the user for the explosion.  The
particles are massless and flow with the fluid, allowing the fluid
dynamics model to capture effects critical for a fireball such as
thermal conductivity and buoyancy.  Some fluid
simulations\cite{Foster:1997:MTM,Stam:1999:SF} model thermal buoyancy
explicitly; in our simulation, thermal buoyancy is a behavior derived
from the governing equations.  For rendering, each particle takes on a
temperature that is interpolated based on its position in the volume.
The particles are rendered as Gaussian blobs with values for red,
green, blue, and opacity.  The color values are based on blackbody
radiation at appropriate wavelengths given the temperature of the
particle\cite{Meyer-Ardent:1984:ICM}.
Figure~\ref{fig:fireball} shows a fireball and corresponding tracer
particles after one second of simulation.  Figure~\ref{fig:corner}
shows a fireball coming around a corner; the hallway is illuminated by
the flames.

The tracer particles couple the appearance of the fireball to the
motion of the fluid, and although heat generated by the initial
explosion is added to the fluid model, any additional heat generated by
post-detonation combustion is ignored.  
Radiative energy released at detonation could also be modeled for
rendering.  Much like the difficulties encountered with rendering the
sun\cite{Preetham:1999:APA}, the high contrast of this effect may
require contrast-reduction techniques such as
LCIS\cite{Tumblin:1999:LAB}.

\begin{figure}[!b]
  \centerline{\includegraphics[width=\columnwidth]{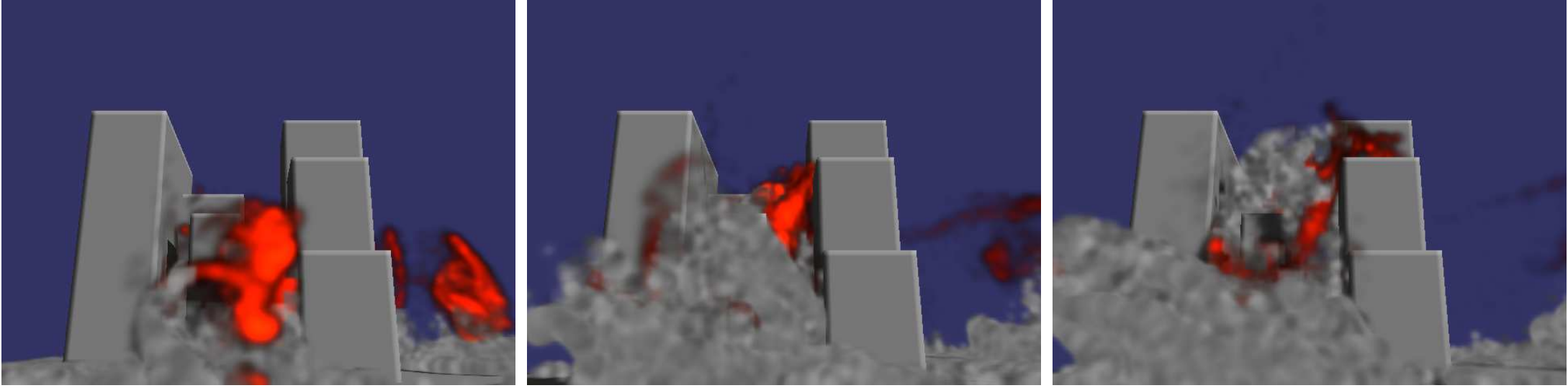}}
  \vspace*{-0.10in} 
  \caption{
    \setlength{\baselineskip}{\captbaselineskip}
    An explosion among buildings.  The images are spaced 667\,ms apart. 
  }\label{fig:city}
  \vspace*{-0.10in}
\end{figure}

The blast wave and other secondary waves create dust clouds by
disturbing fine particles resting on surfaces.  The creation of dust
clouds is difficult to quantify either experimentally or analytically,
so the rate at which the dust becomes airborne is left as a control
for the animator.  Once a dust particle is airborne, its behavior is
dictated by its size.  The smaller it is, the more it is influenced by
drag forces and the less it is influenced by inertial forces.  Smaller
dust particles have lower terminal velocities and exhibit more
Brownian motion.  With the exception of coagulated particles,
experiments reveal that most dust particles are approximately
ellipsoidal with low eccentricity, and the particles do not orient
themselves to the fluid flow\cite{Green:1964:PC}.  The difference in
dynamics between these particles and spherical particles is not that
significant, so we assume dust particles to be spherical.  We
implement dust as metaparticles, each representing a Gaussian density
of homogeneous dust particles.  The dust size for each metaparticle is
chosen according to size distributions from experimental data for
blasted shale\cite{Green:1964:PC}.  The metaparticles travel through
the fluid as if single particles were located at their centers.  Their
variances grow according to the mean Brownian diffusion per unit time.
Figure~\ref{fig:city} shows dust clouds in a city scene.

\section{Results and Discussion}

We ran the system with several scenarios.  The physical constants used
in the simulation were constants for air that were taken from an
engineering handbook\cite{Kuethe:1998:FAB}. Table~\ref{tab:params}
shows the voxel width, timestep, total simulation time, initial volume
of the explosion (proportional to yield), and initial pressure and
temperature of the explosion.  The timesteps $\Delta t$ increase by a
factor of five once the blast wave leaves the volume.

The simulations ran on a single 195\,MHz~R10K processor and used a
$101\times 101\times 101$ volume.  The running times per timestep
varied considerably from several seconds to two minutes because of the
pruning described in Section~\ref{sec:bound}.  For coupling with
fracture, I/O became a major factor because in each iteration the
entire volume was written to disk; however, using better compression
would reduce this expense.  Running times of the simulations varied
from a few hours (Figure~\ref{fig:shaped}) to overnight
(Figures~\ref{fig:wall2d}, \ref{fig:window}, \ref{fig:wallBreak},
and~\ref{fig:grids}) to a few days (Figures~\ref{fig:chimney},
\ref{fig:fireball}, \ref{fig:corner}, and~\ref{fig:nuke}).
\begin{table}[!t]
\begin{center}
\small\textsf{
\label{tab:params}
\begin{tabular}{|l||r|r|r|r|r|r|} \hline
Example          & $h$ & $\Delta t$ & $t_{tot}$ & $V_0$ & $P_0$ & $T_0$ \\
(figure)        & ($\mathrm{m}$) & ($\mathrm{ms}$)   & ($\mathrm{ms}$) & ($\mathrm m^3$) & ($\mathrm{atm}$) & ($\mathrm{K}$) \\
\hline\hline
projectile (\ref{fig:chimney}) & 1.0 & 0.10 & 450 & 73.60 & 1000 & 2900\\
\hline
barrier (\ref{fig:wall2d}) & 0.2 & 0.01 & 25 & 0.52 & 1000 & 2900\\
\hline
shapes (\ref{fig:shaped}) & 1.0 & 0.10 & 30 & 1000.00 & 1000 & 2900\\
\hline
fracture (\ref{fig:window},\ref{fig:wallBreak}) & 0.2 & 0.02 & 20 & 0.52 & 1000 & 2900\\
\hline
fireball (\ref{fig:grids},\ref{fig:fireball}) & 1.0 & 0.10 & 1000 & 65.40 & 1000 & 2900\\
\hline
corner (\ref{fig:corner}) & 1.0 & 0.10 & 10000 & 268.08  & 1000 & 2900\\
\hline
city (\ref{fig:city}) & 1.0 & 0.10 & 5000 & 65.40  & 1000 & 2900\\
\hline
nuclear (\ref{fig:nuke}) & 50.0 & 0.50 & 30000 & $\mathsf{9.1}\!\!\times\!\! \mathsf{10}^\mathsf{7}$  & 345 & $\mathsf{1}\!\!\times\!\!\mathsf{10}^\mathsf{5}$\\
\hline
\end{tabular}}
  \vspace*{-0.10in} 
  \caption{
  \setlength{\baselineskip}{\captbaselineskip}
  Parameters for simulations: voxel width, timestep, total simulation
  time, and initial volume, pressure, and temperature of detonation. 
}
\end{center}
\vspace*{-0.20in}
\end{table}


We use an explicit integration technique to compute the motion of the
pressure wave caused by the detonation.  Despite its magnitude, the
wave does not transport fluid large distances. Previously, fluid
dynamics has been used most often in computer graphics to capture the
effects of macroscale fluid transport where the fluid does move a
significant distance.  Implicit integration techniques with large
timesteps are appropriate for these situations because they achieve
stability by damping high frequencies.  The propagation of the
pressure wave in our stiff equations, however, is characterized by
these high frequencies and it is essential that they not be
artificially damped.  We chose, therefore, to use an explicit
integration technique; however, an implicit integration technique
could be used to simulate the fireball and dust clouds after the blast
wave and the secondary waves have left the volume.  Using an implicit
integration technique in the slow flow regime could allow larger
timesteps and faster execution times.


\pagebreak
We assume that the voxels in the fluid volume are of a size
appropriate for the phenomena that we wish to capture.  In particular,
if solid objects have a dimension smaller than a voxel, then they will
not create a hard boundary that prevents fluid flow.  For example, a
wall that is thinner than a voxel will permit the blast wave to travel
through it because partial volumes do not maintain any geometric
information about the sub-voxel shape of the object.  The difficulty
of a two-way coupling with fracturing objects stems from having to
model subvoxel cracks, which should allow flow to go through.  If
small objects are required, the voxel size could be decreased or
dynamic remeshing techniques could be used to create smaller voxels in
the areas around boundaries.

There are effects from explosions that we have not investigated.
Although smoke is often a visible feature of an explosion that
includes a fireball, we do not have a physically based model for smoke
creation.  Incomplete combustion at lower temperatures results in
smoke, and that observation could be used as a heuristic to determine
where smoke should be created in the fireball and how densely.  Stam
and Fiume used a similar heuristic model\cite{Stam:1995:DFA}.
Textures of objects could be modified to show soot accumulation and
scorching over time.  Dust clouds are created when an object
fractures or pulverizes.  Dust could be introduced into our system
when the finite-element model produces small tetrahedra or when cracks
form.


We made several assumptions in constructing our model of explosions.
Most discounted effects that did not contribute noticeably to the
final rendered images; however, some could produce a noticeable change
in behavior in certain situations and may warrant further
investigation.  We only model the blast wave traveling through air.
However, waves travel through other media, including solid objects,
and complex interface effects occur when a wave travels between two
different media.  For large-scale explosions, meteorological
conditions such as the change in pressure with respect to altitude or
the interface between atmospheric layers (the tropopause) may need to
be considered.

\begin{figure}[!t]
  \centerline{\includegraphics[width=\columnwidth]{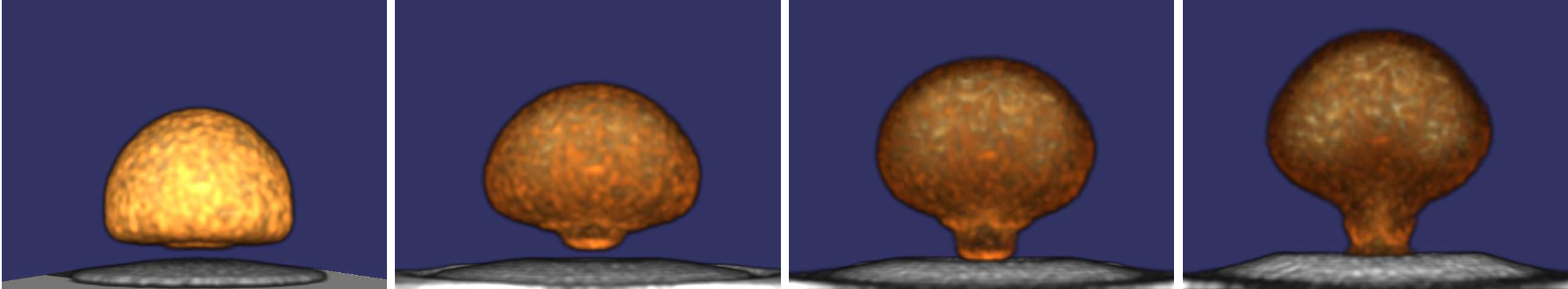}}
  \vspace*{-0.10in} 
  \caption{
    \setlength{\baselineskip}{\captbaselineskip}
    A large-scale high-temperature explosion resembling a nuclear explosion: after 3\,s, 6\,s, 12\,s, 24\,s.
  }\label{fig:nuke}
  \vspace*{-0.10in}
\end{figure}

Our goal in this work has been to create a physically realistic model
of explosions.  However, this model should also lend itself to
creating less realistic effects.  Even though our model does not
incorporate high-temperature effects such as ionization, we can still
obtain interesting results on high-temperature explosions.  The
fireball in Figure~\ref{fig:nuke} resulted from an initial detonation
at $10^5\,K$.  Explosions used in feature films often include far more
dramatic fireballs than would occur in the actual explosions that they
purport to mimic.  By using more tracer particles and adjusting the
rendering parameters of the fireballs, we should be able to reproduce
this effect.  Noise could be added either to the velocity fields or
particle positions post-process to make the explosion look more
turbulent.  Similarly, explosions in space are often portrayed as more
colorful and violent than explosions that occurred outside of the
atmosphere should be.  Imparting an initial outward velocity to the
explosion, turning off gravity, and increasing the thermal buoyancy by
modifying the state equations might create a similar effect.


\section{Acknowledgments}
This project was supported in part by NSF NYI Grant No.~IRI-9457621,
Mitsubishi Electric Research Laboratory, and a Packard Fellowship.
The second author was supported by a Fellowship from the Intel
Foundation.

{\small
  \bibliographystyle{plain-mod}
  \bibliography{explosions}
}




\end{document}